# Quantum Telecomputation

Lov K. Grover, *lkgrover@bell-labs.com*

## Summary

Quantum mechanics permits certain kinds of non-local effects. This paper demonstrates how these can be used for distributed computation with minimal communication between various processors. The problem considered is that of estimating the mean of $N$ items to a precision $\varepsilon$, i.e. if there are $N$ numbers that lie in the range $-1$ to 1, the estimate, $\mu_e$, should be such that, with a large probability, the actual mean of the $N$ numbers lies in the range $\mu_e \pm O(\varepsilon)$.

(i) Any classical algorithm for this will need at least $\Omega\left(\frac{1}{\varepsilon^2}\right)$ samples. By using quantum mechanical effects this paper first presents a serial algorithm that requires only $O\left(\frac{1}{|\varepsilon|^{1.5}}\right)$ time steps.

(ii) Using a quantum mechanical system consisting of $\eta$ coupled EPR particles, the computation time in (i) above, can be reduced by a factor of $O(\eta)$. The $\eta$ coupled EPR particles can be remotely distributed and work independently. Each of them has just to transmit one bit of classical information to a central location.

**0.0 Background** Paradoxical effects with coupled particles are well known in quantum mechanics. There are three well known examples of this. The first is the original EPR paradox. This was proposed by Einstein (a famous skeptic of quantum mechanics) and others in 1935 and was meant to deal a deathblow to quantum mechanics. In this, two particles that were once in close proximity and then separated, somehow stay coupled. Operations carried out on one particle instantaneously affect the other particle. This seemed to violate the constraint that physical influences cannot propagate faster than the speed of light. However, it was soon realized that this could not be used to transmit any information and thus did not violate any physical law.

Also in 1935, Schrodinger devised the paradox of the Schrodinger cat. In this, a macroscopic object (a cat) is in a superposition of two states, one in which it is dead and the other in which it is alive, and the question is regarding what *really* happens when it collapses into a well defined state. The results of this paradox though puzzling did not violate any physical law. This paper uses a Schrodinger cat state, i.e. a large system in a superposition of two states. The computational power of the system is shown to be proportional to the number of particles in the system.

The third paradoxical effect of coupled particle states is quantum teleportation. In this a quantum state can be transmitted between two parties that share an EPR pair, merely by transmitting two bits of classical information and carrying out some processing at each end. This too, though seemingly paradoxical, was after some analysis, seen not to violate any physical law since the speed at which any information is transmitted is limited by the time it takes for the classical information to get through.

**0.1 This paper** The reason for these paradoxes is that quantum mechanics permits non-local effects. This paper uses such non-local effects to carry out quantum computing when the processors are at remote locations. A group of $\eta$ particles is placed in just two composite states. This is called a cat state by analogy with Schrodinger's cat, i.e a system of many particles that is in two possible states. These $\eta$ particles are separated to different locations; each particle can then independently carry out its computations and convey the necessary information to the base station, just by transmitting one bit of classical information. The overall computation time is faster by a factor of $O(\eta)$.

The other contribution of this paper is a novel quantum mechanical algorithm for an important computer science problem which is faster than any classical algorithm. This is then adapted to distributed computation. The computer science problem is that of estimating the mean of $N$ items to a precision $\varepsilon$, i.e. if there are $N$ numbers that lie in the range $-1$ to 1, the estimate $\mu_e$ should be such that, with a large probability, the actual mean of the $N$ numbers lies in the range $\mu_e \pm O(\varepsilon)$. In case $\nu$ random samples are picked classically, it follows from the central limit theorem that the mean of the $\nu$ samples will be a gaussian distribution centered about the actual mean with a standard deviation of $O\left(\frac{1}{\sqrt{\nu}}\right)$. Therefore, with a probability approaching unity, the estimated mean lies within $O\left(\frac{1}{\sqrt{\nu}}\right)$ of the true mean. In order for the uncertainty due to this variation to become less than $\varepsilon$, $O\left(\frac{1}{\sqrt{\nu}}\right)$ must be smaller than $\varepsilon$, i.e. $\nu$ must be greater than $\Omega\left(\frac{1}{\varepsilon^2}\right)$. Therefore $\Omega\left(\frac{1}{\varepsilon^2}\right)$ samples are needed to estimate the mean with a precision of $\varepsilon$. Quantum mechanical algorithms can simultaneously explore multiple states, it is not clear what (if any) the lower bound for estimating the mean is. This paper gives an algorithm for estimating the mean in $O\left(\frac{1}{|\varepsilon|^{1.5}}\right)$ elementary quantum mechanical steps.

## 1. Quantum mechanical algorithms

Quantum computation and quantum teleportation are two exciting new developments in quantum physics. Quantum computation allows an exponential amount of computing to simultaneously take place in a single piece of hardware. Quantum teleportation allows the transmission of a complex state with arbitrary precision (provided the proper initial conditions are set up) just by transmitting a few bits of classical information. This paper shows that significant quantum computation can take place with the processors being remotely located with each processor needing to transmit at most one bit of classical information.

A good starting point to think of quantum mechanical algorithms is probabilistic algorithms [Model93] (e.g. simulated annealing). In these algorithms, instead of the system being in a specified state, it is in a distribution over various states with a certain probability of being in each state. At each step, there is a certain probability of making a transition from one state to another. The evolution of the system is obtained by premultiplying this probability vector (that describes the distribution of probabilities over various states) by a state transition matrix. Knowing the initial distribution and the state transition matrix, it is possible, in principle, to calculate the distribution at any later time.

Just like classical probabilistic algorithms, quantum mechanical algorithms work with a probability distribution over various states. However, unlike classical systems, the probability vector does not completely describe the system. In order to completely describe the system, we need the *amplitude* in each state which is a complex number. The evolution of the system is obtained by premultiplying this amplitude vector (that describes the distribution of amplitudes over various states) by a transition matrix, the entries of which are complex in general. The probabilities in any state are given by the square of the absolute values of the amplitude in that state. It can be shown that in order to conserve probabilities, the state transition matrix has to be unitary [Model93].

The machinery of quantum mechanical algorithms is illustrated by discussing two operations that are needed in the algorithm of this paper. The first is the creation of a configuration in which the amplitude of the system being in any of the $2^n$ basic states of the system is equal, this is generalized to give the Walsh-Hadamard (WH) transformation; the second is the selective rotation of the phase of different states.

A basic operation in quantum computing is that of a "fair coin flip" performed on a single bit whose states are 0 and 1 [Search96]. This operation is represented by the following matrix: $M = \frac{1}{\sqrt{2}}\begin{bmatrix} 1 & 1 \\ 1 & -1 \end{bmatrix}$. A bit in the state 0 is transformed into a superposition in the two states: $\left(\frac{1}{\sqrt{2}}, \frac{1}{\sqrt{2}}\right)$. Similarly a bit in the state 1 is transformed into $\left(\frac{1}{\sqrt{2}}, -\frac{1}{\sqrt{2}}\right)$, i.e. the magnitude of the amplitude in each state is $\frac{1}{\sqrt{2}}$ but the *phase* of the amplitude in the state 1 is inverted. The phase does not have an analog in classical probabilistic algorithms. It comes about in quantum mechanics since the amplitudes are in general complex. This results in *interference* of different possibilities as in wave mechanics and is what distinguishes quantum mechanical systems from classical probabilistic systems.

In a system in which the states are described by *n* bits (the system has $2^n$ possible states) we can perform the transformation $M$ on each bit independently in sequence thus changing the state of the system. The state transition matrix representing this operation will be of dimension $2^n \times 2^n$. In case the initial configuration was the configuration with all *n* bits in the 0 state, the resultant configuration will have an identical amplitude in each of the $2^n$ states. This is a way of creating a distribution with the same amplitude in all $2^n$ states.

Next consider the case when the starting state is another one of the $2^n$ states, i.e. a state described by an $n$ bit binary string with some 0s and some 1s. The result of performing the transformation $M$ on each bit will be a superposition of states described by all possible $n$ bit binary strings with amplitude of each state having an equal magnitude and sign either + or −. To deduce the sign, observe from the definition of $M$, i.e. $M = \frac{1}{\sqrt{2}}\begin{bmatrix} 1 & 1 \\ 1 & -1 \end{bmatrix}$, that the sign of the resulting state is changed only when a bit that was initially a 1 remains a 1 after the transformation. Hence if $\bar{x}$ be the $n$-bit binary string describing the starting state and $\bar{y}$ the $n$-bit binary string describing the resulting string, the sign of the amplitude of $\bar{y}$ is determined by the parity of the bitwise dot product of $\bar{x}$ and $\bar{y}$, i.e. $(-1)^{\bar{x} \cdot \bar{y}}$. This transformation is referred to as the Walsh-Hadamard (WH) transformation [Transform92]. It is one of the features that makes quantum mechanical algorithms more powerful than classical algorithms and, either this or a closely related transform called the Fourier Transform, forms the basis for most significant quantum mechanical algorithms.

The other transformation that we need is the selective rotation of the phase of the amplitude in certain states. The transformation matrix describing this for a 2 state system is of the form: $\begin{bmatrix} e^{i\phi_1} & 0 \\ 0 & e^{i\phi_2} \end{bmatrix}$ where $i = \sqrt{-1}$ and $\phi_1, \phi_2$ are arbitrary real numbers. Unlike the WH transformation and other state transition matrices, the probability in each state stays the same since the square of the absolute value of the amplitude in each state stays the same. In the following algorithm, the application of this phase rotation transform will be written as: *the phase of the $j^{th}$ state is rotated by $\phi_j$ radians.* This conditional phase shift is the most difficult to implement, it requires two qubits and is responsible for *quantum entanglement* in the system - this is what leads to non-local effects.

## 2.0 Telecomputation
Distributed computing is the problem where multiple processors are remotely distributed and there is a cost associated with communication. The issue in distributed computing is how to optimally divide the problem among the various processors. The algorithms are, in general, different depending on the specific nuance of the problem, e.g. how severe is the cost of communication as compared to computing, how much of the data does each processor have access to, etc. [Qntcmm97] has very recently (independently) shown that quantum entanglement can indeed aid in multi-party communication. This paper shows that a certain important computer science problem can be solved efficiently with distributed quantum processors. The problem is that of finding the average of $N$ real numbers in the range $[-1, 1]$ to a specified precision. The following is the organization of the rest of the paper:

(i) Sections 2.1 & 2.2 consider the problem where the number of processors exactly equal the number of pieces of data. Each processor has access to just one piece of data. The processors can all be at different locations and carry out their computations independently. At the end, each processor has just to transmit one bit of classical information to a base location. This algorithm is similar to a technique that was independently invented in a very different context, i.e. frequency standards & spectroscopy [Freq96] as Artur Ekert pointed out.

(ii) Sections 3 & 4 consider the problem where the number of processors is only a fraction of the number of pieces of data. For this, each processor needs to carry out several computations before the results of all the processors can be combined. There are thus two parts:

 (a) First is a serial $O\left(\frac{1}{|\varepsilon|^{1.5}}\right)$ step quantum mechanical algorithm for estimating the mean of $N$ numbers in the range $[-1, 1]$ (any classical algorithm will take at least $\Omega\left(\frac{1}{|\varepsilon|^2}\right)$ steps). (Sections 3.3 & 3.4).

 (b) Next, it is shown that with $\eta$ appropriately initialized quantum processors at remote locations, it is possible to carry out the overall computation $O(\eta)$ times faster. Each processor operates independently; at the end it transmits just one bit of classical information to the base location. (Sections 4.0, 4.1 & 4.2).

## 2.1 First Problem

Consider the following problem: there are $N = 2^n$ states $S_0, S_1, ... S_{N-1}$ each of which has an associated scalar value $v_j$ in the range $[-1, 1]$. Assume that there is a quantum processing element for each state (since each quantum processor is usually just a single particle, this is not extravagant in terms of the hardware required). The mean $\mu$ is defined by $\mu \equiv \frac{1}{N} \sum_{j=0}^{N-1} v_j$. Let $|\mu|$ be known to be $O(\theta^2)$ (where $\theta$ is a small quantity). The object is to estimate this mean with a precision of $O(\theta^2)$, i.e. the estimate $\mu_e$ should be such that, with a large probability, the actual mean of the $N$ numbers lies in the range $\mu_e \pm O(\theta^2)$.

The problem when the order of the mean is unknown can be solved in a logarithmic number of repetitions of the algorithm used to solve the above problem. The order of the mean can be estimated by initially assuming the mean ($\mu$) to be bounded by a large quantity ($\theta$ is large), estimating the mean, and keep scaling down $\theta$ in each iteration sequence until $\mu_e$ becomes non-trivial. For example, assume $\mu$ to be $2 \times 10^{-8}$. Initially $\theta$ would be set to a large value, say 0.5, and the mean estimated with a precision of $O(\theta^2)$, say $0.1\theta^2$ which is $2.5 \times 10^{-2}$; the result will therefore be somewhere in the range $\pm 0.1\theta^2$, i.e. $\pm 2.5 \times 10^{-2}$. Next $\theta$ is lowered by a constant factor (say 1.5) and the procedure repeated. The result will again be in the range $\pm 0.1\theta^2$ (with the reduced value of $\theta$). $\theta$ would be repeatedly lowered until the absolute value of the result becomes greater than $0.1\theta^2$. This happens after 18 repetitions, when $\theta$ has fallen to $3.35 \times 10^{-4}$. A similar procedure is also described in [Median96].

## 2.2 Algorithm

The following steps estimate $\mu$ with a precision of $O(\theta^2)$ ($\theta$ was just defined above in section 2.1, by the statement that $\mu$ *be known to be* $O(\theta^2)$). Angular brackets represent averages, e.g. $\langle v \rangle \equiv \frac{\sum v_j}{N}$.

Consider the following quantum state vector of $N$ particles: $(|0\rangle_0 |0\rangle_1 ... |0\rangle_{N-1} + |1\rangle_0 |1\rangle_1 ... |1\rangle_{N-1})$, this is referred to as *an $N$ particle EPR system* (as mentioned earlier, this is called a cat state by analogy with Schrodinger's cat, i.e a system of many particles that is in two overall states). The state has to be prepared when the $N$ particles are together. However once the state is prepared, they can be separated and taken to distant locations. The phase of the $|0\rangle$ state of the $j^{th}$ particle is rotated by $\frac{v_j}{N\theta^2}$. It follows that the state vector of the system after this operation becomes: $\left(|0\rangle_0 |0\rangle_1 ... |0\rangle_{N-1} \exp\left(i\frac{\langle v \rangle}{\theta^2}\right) + |1\rangle_0 |1\rangle_1 ... |1\rangle_{N-1}\right)$. Note that the particles are still physically separated. Next apply the operation $M$, as described in section 1.1, to each of the $(N-1)$ particles: $(1, 2 ... N-1)$. The state vector becomes proportional to: $\left(|0\rangle_0 (|0\rangle + |1\rangle)_1 ... (|0\rangle + |1\rangle)_{N-1} \exp\left(i\frac{\langle v \rangle}{\theta^2}\right) + |1\rangle_0 (|0\rangle - |1\rangle)_1 ... (|0\rangle - |1\rangle)_{N-1}\right)$.

Now if the state of each of the $(N-1)$ particles: $(1, 2 ... (N-1))$ is measured, it follows that the state of the $0^{th}$ particle becomes $\left(|0\rangle_0 \exp\left(i\frac{\langle v \rangle}{\theta^2}\right) \pm |1\rangle_0\right)$ where the sign (plus or minus) depends on the results of the $(N-1)$ measurements. In case an even number of the $(N-1)$ measurements be 1's, the sign is positive, if an odd number be 1's, the sign is negative. Therefore if the results of all these measurements be transmitted to the location of the $0^{th}$ particle, the state of the $0^{th}$ particle becomes precisely known i.e. we have a single binary state particle where the phase of one of the states is rotated by a constant amount proportional to $\langle v \rangle$.

A constant phase shift can be estimated by an interference type experiment. For example consider the matrix $M = \frac{1}{\sqrt{2}}\begin{bmatrix} 1 & 1 \\ 1 & -1 \end{bmatrix}$ as discussed in section 1.1. When $M$ acts on a state vector proportional to $\begin{bmatrix} 1 \\ \exp(i\Theta) \end{bmatrix}$, the resulting state vector is proportional to $\begin{bmatrix} 1 + \exp(i\Theta) \\ 1 - \exp(i\Theta) \end{bmatrix}$. The ratio of the probabilities in the two states is $\left(\cos\frac{\Theta}{2}\right)^2 : \left(\sin\frac{\Theta}{2}\right)^2$. Therefore if the state of the system now be measured, it is possible to estimate the phase $\Theta$. By repeating this experiment $\alpha$ times, the phase $\Theta$ can be estimated with a precision of $\frac{1}{\sqrt{\alpha}}$.

The above applies both to the case where the data is localized and also when it is distributed. In the former case it gives a parallel algorithm for adding $N$ numbers by doing $N$ simultaneous unitary transforms on $N$ particles and then calculating the XOR of $N$ bits. Alternatively, in the more interesting case, if the data is distributed, the benefit is that instead of transmitting all of the data to the base location, we have to transmit just one bit of classical data from each location to the base location. Since the XOR is commutative and associative, it follows that in case some of the data were located together, then instead of transmitting one bit for each datapoint, we could XOR the local results and just transmit the result of the local XOR.

### 3.1 Second Problem
Consider the problem discussed in section 2.1, with the difference that the number of processors, $\eta$, is only a small fraction of the number of datapoints $N$ (previously the number of processors was equal to the number of datapoints). Sections 3.3 & 3.4 give an $O\left(\frac{1}{\theta^3}\right)$ step serial algorithm using just one quantum processor. This corresponds to the situation discussed in section 0.1 with $\theta^2 = \varepsilon$; i.e. in the terminology of section 0.1, the complexity is $O\left(\frac{1}{|\varepsilon|^{1.5}}\right)$. Sections 4.0, 4.1 & 4.2 consider the multiprocessor case.

### 3.2 Outline
Given a particle in a state $|0\rangle$, the following steps (described in section 3.3) rotate the phase to produce the state $|0\rangle\exp(i\phi)$. Here $\phi$ is proportional to $\langle v \rangle + O(Max(v_j^3))$. The second term, i.e. $O(Max(v_j^3))$ is an error term. In order to reduce its relative contribution, scale the values of each state by $\theta$ and define $x_j \equiv \theta v_j$. Hence each $x_j$ is $O(\theta)$ and $\langle x \rangle$ is $O(\theta^3)$ and so the relative value of the error gets reduced. The following steps estimate $\langle x \rangle$ with a precision of $O(\theta^3)$, thus leading to an estimate of the mean, $\mu$, with a precision of $O(\theta^2)$ ((i) & (iii) are the same steps as already carried out in section 2.2)

(i)   Starting from $(|0\rangle + |1\rangle)$, the system is put into the state $(|0\rangle\exp(i\phi) + |1\rangle)$, where $\phi$ is proportional to $\langle x \rangle$.
(ii)  This relative phase, $\phi$, can be measured by making it apparent in the relative probabilities of two states in an interference type experiment and doing a statistical sampling of several such systems.
(iii) This idea can be extended to a "cat state" (an $\eta$ particle EPR system with the $\eta$ particles at remote locations) where all the phase rotation occurs in the same term and hence adds.

### 3.3 Algorithm
This section discusses the serial algorithm. Given a particle in a state $|0\rangle$, the following steps rotate the phase to produce the state $|0\rangle\exp(i\phi)$. Here $\phi$ is proportional to $\langle x \rangle$ and is of magnitude $O(1)$ (as men-tioned earlier, $\langle x \rangle$ is $O(\theta^3)$ and so the proportionality constant relating $\phi$ to $\langle x \rangle$ will have to be $O\left(\frac{1}{\theta^3}\right)$. This result is achieved with a probability of $O(\theta^4)$.

The algorithm, though stated below for a single state $|0\rangle$, can be easily extended to the case where the phase of only one of the many states of the system needs to be rotated, i.e. starting from $|0\rangle + |1\rangle$, produce the state $|0\rangle \exp(i\phi) + |1\rangle$. This is achieved by carrying out the operations conditioned on the fact that the system is in state $|0\rangle$. It also applies to the case where the system is arbitrarily coupled to the environment, i.e. starting from the state $|0\rangle|e\rangle + |1\rangle|f\rangle$, produce the state $|0\rangle|e\rangle \exp(i\phi) + |1\rangle|f\rangle$.

(i) Starting from the $|0\rangle$ state (the state the phase of which needs to be rotated), apply the WH transform to initialize the system so that there is the same amplitude in each of the $N = 2^n$ states (as described in section 1).

*Amplitude in every state:* $\frac{1}{\sqrt{N}}$

(ii) In case the system is in the $j^{th}$ state, rotate the phase by $\gamma_j$ where $\sin \gamma_j = x_j$ (since $x_j$ is small, $\gamma_j \approx x_j$).

*Amplitude in the $j^{th}$ state:* $\frac{1}{\sqrt{N}}\left(\sqrt{1 - x_j^2} + ix_j\right)$

(iii) Apply the WH Transform as discussed in section 1.1: $W_{pq} = 2^{-n/2}(-1)^{\bar{p} \cdot \bar{q}}$, where $\bar{p}$ is the binary representation of $p$, and $\bar{p} \cdot \bar{q}$ is the bitwise dot product of the $n$ bit strings $\bar{p}$ & $\bar{q}$.

*Amplitude in the $0^{th}$ state:* $\left(\langle\sqrt{1 - x^2}\rangle + i\langle x \rangle\right)$

(iv) In case the system is in the $0^{th}$ state, rotate the phase by $\pi$ radians.

(v) Apply the WH Transform.

*Amplitude in the $j^{th}$ state:* $\frac{1}{\sqrt{N}}\left(\sqrt{1 - x_j^2} - 2\langle\sqrt{1 - x^2}\rangle\right) + \frac{i}{\sqrt{N}}(x_j - 2\langle x \rangle)$

(vi) In case the system is in the $j^{th}$ state, rotate the phase by $\gamma_j$ where $\sin \gamma_j = x_j$ (since $x_j$ is small, $\gamma_j \approx x_j$).

*Amplitude in the $j^{th}$ state:* $\left(-\frac{1}{\sqrt{N}} - \frac{2i\langle x \rangle}{\sqrt{N}}\right) + \frac{2}{\sqrt{N}}\left(\sqrt{1 - x_j^2} - \langle\sqrt{1 - x^2}\rangle\right) + \left(\frac{1}{\sqrt{N}}O(\theta^4) + \frac{i}{\sqrt{N}}O(\theta^3)\right)$

(vii) Apply the WH Transform.

*Amplitude in the $0^{th}$ state:* $(-1 - 2i\langle x \rangle + (O(\theta^4) + iO(\theta^3)))$

(viii) Make an observation that *measures* whether or not the system is in the $0^{th}$ state. With a probability of $(1 - O(\theta^4))$, the system will indeed be in the $0^{th}$ state with an amplitude proportional to $(-1 - 2i\langle x \rangle + (O(\theta^3) + iO(\theta^3)))$, in case it is not in the $0^{th}$ state, stop and start from the beginning.

The net effect of the algorithm is to rotate the phase of the amplitude of the $|0\rangle$ state by approximately $2\langle x \rangle \pm O(\theta^3)$. If this procedure be repeated $r$ times, the phase gets rotated by approximately $2r\langle x \rangle$. Since $\langle x \rangle$ is $O(\theta^3)$, it follows that if the number of repetitions, $r$, be chosen to be $O\left(\frac{1}{\theta^3}\right)$, the phase gets rotated by a finite amount proportional to $\langle x \rangle$. As described earlier, in section 2.2, a finite phase shift can be estimated by interacting with another state in the reference phase - note that this will require $O\left(\frac{\alpha}{\theta^3}\right)$ repetitions of the sequence (i),...(viii) in the algorithm since, as just mentioned, it requires $O\left(\frac{1}{\theta^3}\right)$ repetitions of the sequence (i)(ii)...(viii) to rotate the phase by O(1). Since $\alpha$ is $O(1)$, the total number of steps required will thus be $O\left(\frac{1}{\theta^3}\right)$ which is $O\left(\frac{1}{|\epsilon|^{1.5}}\right)$. In the measurement of step (viii), the probability of the system *not* being in the $0^{th}$ state is $O(\theta^4)$, it follows that in $O\left(\frac{1}{\theta^3}\right)$ repetitions of the procedure there will only be an $O(\theta)$ probability of observing the system in a state other than the $0^{th}$ state.

Hence $\langle x \rangle$ can be estimated with a precision of $O(\theta^3)$; $\mu$, which is $\frac{\langle x \rangle}{\theta}$, is thus estimated with a precision of $O(\theta^2)$ in accordance with the problem specification of section 2.1.

### 3.4 How does the algorithm work?
The iteration sequence (i), (ii)...(viii) in section 3.3, rotates the phase of the state $|0\rangle$ by an amount proportional to $\langle x \rangle$. In order to see this, observe:

(ii)     Denote the amplitude after step (ii) in state $j$ by $a_j$, i.e. $a_j \equiv \frac{1}{\sqrt{N}}\left(\sqrt{1-x_j^2} + ix_j\right)$. The state vector after step (ii) is thus $(a_0, a_1, ..., a_{N-1})$.

(iii)     Denote the amplitude in state $j$, after the WH transform of step (iii), by $w_j$. The state vector after step (iii) is thus $(w_0, w_1...w_{N-1})$. It follows by the definition of the WH transform in section 1.1, that $w_0 = \frac{1}{\sqrt{N}}\sum_{j=0}^{N-1} a_j$, which by using the value for $a_j$ from (ii) above, becomes $w_0 = \left(\langle\sqrt{1-x^2}\rangle + i\langle x \rangle\right)$.

As can be seen by the expression for $w_0$, the phase after (iii) in the $0^{th}$ state is proportional to $\langle x \rangle$. However, if we measure whether or not the system is in the $0^{th}$ state, there will be a relatively high probability of it collapsing to a non-zero state. The following steps (iv)..(viii) serve to reduce this probability.

(iv)     By inverting the phase of the $0^{th}$ state (in step (iv)), the state vector becomes: $(-w_0, w_1...w_{N-1})$.

(v)     The state vector after step (iv) can be written as:
$(-w_0, w_1...w_{N-1}) = (-2w_0, 0, 0...0) + (w_0, w_1...w_{N-1})$. Since the WH transform is its own inverse, it follows that the WH transform of $(w_0, w_1...w_{N-1})$ is $(a_0, a_1, ..., a_{N-1})$. By the definition of the WH transform in section 1.1, the WH transform of $(-2w_0, 0, 0...0)$ is $\left(-\frac{2w_0}{\sqrt{N}}, -\frac{2w_0}{\sqrt{N}}, ... -\frac{2w_0}{\sqrt{N}}\right)$; substituting the value of $w_0$ from (iii), it follows that the total amplitude in state $j$ after step (v) is:

$\frac{1}{\sqrt{N}}\left(\sqrt{1-x_j^2} - 2\langle\sqrt{1-x^2}\rangle\right) + \frac{i}{\sqrt{N}}(x_j - 2\langle x \rangle)$.

(vi)     The amplitude in state $j$, after step (v), as derived above, can be written as:

$$\frac{1}{\sqrt{N}}\left(-\sqrt{1-x_j^2}+ix_j\right)-\frac{2i\langle x\rangle}{\sqrt{N}}+\frac{2}{\sqrt{N}}\left(\sqrt{1-x_j^2}-\langle\sqrt{1-x^2}\rangle\right).$$ The first two terms are the dominant ones, the last term (in parentheses) contributes the *error* terms. When the phase is rotated by $x_j$, it follows that since each $x_j$ is $O(\theta)$ & $\langle x\rangle$ is $O(\theta^3)$, the above becomes:

$$-\frac{1}{\sqrt{N}}-\left(\frac{2i\langle x\rangle}{\sqrt{N}}(1-O(\theta^2))+\frac{1}{\sqrt{N}}O(\theta^4)\right)+\frac{2}{\sqrt{N}}\left(\sqrt{1-x_j^2}-\langle\sqrt{1-x^2}\rangle\right)+\left(\frac{1}{\sqrt{N}}O(\theta^4)+\frac{i}{\sqrt{N}}O(\theta^3)\right)$$

This may be written as: $\left(-\frac{1}{\sqrt{N}}-\frac{2i\langle x\rangle}{\sqrt{N}}\right)+\frac{2}{\sqrt{N}}\left(\sqrt{1-x_j^2}-\langle\sqrt{1-x^2}\rangle\right)+\left(\frac{1}{\sqrt{N}}O(\theta^4)+\frac{i}{\sqrt{N}}O(\theta^3)\right)$

(vii) When the WH transform is taken, as mentioned in (iii) above, the amplitude in the $0^{th}$ state is $w_0=\frac{1}{\sqrt{N}}\sum_{j=0}^{N-1}a_j$. The contribution from the term $\frac{2}{\sqrt{N}}\left(\sqrt{1-x_j^2}-\langle\sqrt{1-x^2}\rangle\right)$ is hence zero. Therefore the amplitude in the $0^{th}$ state is $(-1-2i\langle x\rangle+(O(\theta^4)+iO(\theta^3)))$. Therefore by precisely estimating the phase of the state as described at the end of section 2.2, $\langle x\rangle$ can be estimated with a precision of $O(\theta^3)$.

(viii) The probability of the system not being in the $0^{th}$ state is obtained from the expression for the amplitude in (vi). It is easily deduced that the contribution of the term $\left(-\frac{1}{\sqrt{N}}-\frac{2i\langle x\rangle}{\sqrt{N}}\right)$ to any state $j, j\neq 0$ is zero. By using the fact that each $x_j$ is $O(\theta)$, it follows that the probability due to the rest of the terms is $O(\theta^4)$.

## 4.0 Parallel Computing

Section 2.2 discussed a simple case of quantum mechanical parallel computing where each processor had just to carry out a single operation. In general each processor has to carry out a finite number of operations and then share its result(s) with other processors. A problem with quantum mechanical systems is that it is not possible to make a copy of a quantum state.

The main issue in parallel computing, whether classical or quantum mechanical, is regarding how to partition the problem among different processors. If the problem of estimating the mean was being solved classically, the partitioning would be obvious, i.e. in case $O\left(\frac{1}{\varepsilon^2}\right)$ samples have to be generated in all, have each of the $\eta$ processors generate $O\left(\frac{1}{\eta\varepsilon^2}\right)$ samples. This immediately gives a speed-up by $O(\eta)$. This procedure does not work with the quantum mechanical algorithm of section 3.3 because each of the processors only calculates the mean with a finite precision. Whereas the classical processors can calculate the mean of a certain fraction of randomly chosen samples exactly, the quantum mechanical algorithm of section 3.3 calculates the mean of all of the datapoints, but only with a certain precision. For example, if we allow each of the $\eta$ processors $O\left(\frac{1}{\eta\varepsilon^{1.5}}\right)$ steps, and then classically calculate the mean of the $\eta$ values, it is easily shown that the final result is only precise to $O(\eta^{0.166...}\varepsilon)$. If the final result has to be precise to $O(\varepsilon)$, each of the processors has to go through $O\left(\frac{1}{\eta^{0.75}\varepsilon^{1.5}}\right)$ steps, therefore this only gives a speed-up of $O(\eta^{0.75})$.

Another possibility is to take the $\eta$ quantum mechanical binary state systems obtained after the algorithm of sec-tion 3.3 and carry out suitable unitary & projection operations to try to get a single binary state system whose phase is rotated by a larger amount. The way such an approach might work is the following. Consider two independent quantum mechanical systems which have been through the procedure described in section 3.3. The phase of state $|0\rangle$ of both systems is rotated by $\phi$. The composite system is thus $(|0\rangle_1 \exp(j\phi) + |1\rangle_1)(|0\rangle_2 \exp(j\phi) + |1\rangle_2)$. This may be written as: $(|00\rangle \exp(2j\phi) + |01\rangle \exp(j\phi) + |10\rangle \exp(j\phi) + |11\rangle)$. If this is passed through a CONTROLLED-NOT gate with the first bit as the control, the new state becomes: $(|00\rangle \exp(2j\phi) + |01\rangle \exp(j\phi) + |11\rangle \exp(j\phi) + |10\rangle)$. If the second bit is now measured, then with a 0.5 probability it will be a 0, in which case the first bit will be in the state $(|0\rangle \exp(2j\phi) + |1\rangle)$, i.e. the angle by which the second state is rotated has got doubled. Therefore if we have $\eta$ such quantum mechanical systems with the phase of the second state of each of them rotated by $\phi$, then after repeating the above process on $\frac{\eta}{2}$ pairs, we obtain approximately $\frac{\eta}{4}$ binary state systems where the phase of the second state is rotated by $2\phi$. Assuming $\eta$ to be a power of 2, it follows that after $O(\log \eta)$ iterations, we obtain O(1) systems where the phase of the second state is rotated by $O(\sqrt{\eta}\phi)$. This process thus gives a speed up of only $O(\sqrt{\eta})$. Section 4.2 describes a parallel algorithm with speed up of $O(\eta)$.

### 4.1 Distributed computing

Distributed computing is the case of parallel computing where the processors are remotely distributed and there is a cost associated with communication between them. We assume that all processors have access to either all the data or a representative portion of the data - this would be the case if the data were being computed by a simply specified rule. The algorithm easily extends to the case where different processors have access to different portions of the data, but we postpone discussing that to another occasion.

The classical solution to this problem is immediate if each of the processors is able to communicate the result of its computation to a central location. It is then the same as the one discussed at the beginning of the previous section, i.e. have each processor generate $O\left(\frac{1}{\eta \varepsilon^2}\right)$ samples.

The obvious quantum mechanical solutions to this are the same as the two discussed in section 4.0. In the first suggestion, each processor has to communicate the full result of its computation to the central location, the speed up is $O(\eta^{0.75})$. In the second algorithm (where multiple systems were combined to give another system with a larger phase rotation), the speed up is only $O(\eta^{0.5})$; however it is now possible to use quantum teleportation [Teleport93] to transfer the processed quantum system to the central location using just two bits of classical information instead of the full result of the computation. This leads us into the final algorithm where the speed up is $O(\eta)$, also each processor has just to transmit one bit of classical information to a central location.

### 4.2 Distributed implementation of algorithm of section 3.3

Assume that all processors have access to either all the data or a representative portion of the data. As mentioned before in section 3.3, since the result of the algorithm does not depend on how the state of the particle being rotated is coupled to other particles, the algorithm is easily extended to the case where the phase of only one of the many states of the system needs to be rotated, and the system is arbitrarily coupled to the environment, i.e. starting from the state $|0\rangle|e\rangle + |1\rangle|f\rangle$, produce the state $|0\rangle|e\rangle \exp(i\phi) + |1\rangle|f\rangle$.

As in section 2.2, consider the following quantum state of $\eta$ particles: $(|0\rangle_0 |0\rangle_1 \ldots |0\rangle_{\eta-1} + |1\rangle_0 |1\rangle_1 \ldots |1\rangle_{\eta-1})$. The algorithm of section 3.3 can be applied to each of the $\eta$ particles independently, to rotate the phase of the $|0\rangle$ state of each particle by $\phi$. The state of the system after this operation becomes: $(|0\rangle_0 |0\rangle_1 \ldots |0\rangle_{\eta-1} \exp(i\eta\phi) + |1\rangle_0 |1\rangle_1 \ldots |1\rangle_{\eta-1})$. Note that the particles are still physically separated. Next apply the operation $M$ (as described in section 1.1) to each of the $(\eta - 1)$ particles: $1, 2 \ldots \eta - 1$. The state now

becomes proportional to: $(|0\rangle_0(|0\rangle + |1\rangle)_1...(|0\rangle + |1\rangle)_{\eta-1}\exp(i\eta\phi) + |1\rangle_0(|0\rangle - |1\rangle)_1...(|0\rangle - |1\rangle)_{\eta-1})$. Now if the state of the $(\eta-1)$ particles: $(1, 2...\eta-1)$ are measured, it is easily seen that the state of the $0^{th}$ particle becomes $(|0\rangle_0\exp(i\eta\phi)\pm|1\rangle_0)$ where the sign (plus or minus) depends on the results of the $(\eta-1)$ measurements. In case an even number of the $(\eta-1)$ measurements be 1's, the sign is positive, if an odd number be 1's, the sign is negative Therefore if the results of all these measurements be transmitted to the location of the $0^{th}$ particle, the state of the $0^{th}$ particle becomes precisely known and we have accomplished our objective, i.e. of using $\eta$ remotely located binary state particles, each of which have the phases of one of their states rotated by $\phi$, to obtain a single binary state particle in which the phase of one of the states is rotated by $\eta\phi$.

The additional constraint in working with an $\eta$ particle system is that the total probability of obtaining a non-zero result in step (viii) of the algorithm be negligible for all $\eta$ particles, i.e. $\eta \times O(\theta^4) \times O\left(\frac{1}{\theta^3}\right)$ is O(1).

Therefore $\eta$ should be less than $O\left(\frac{1}{\theta}\right)$. The number of processors has to be less than $O\left(\frac{1}{\theta}\right)$ or equivalently $O\left(\frac{1}{\langle x\rangle^{0.5}}\right)$, this gives a bound on the speedup obtainable.

## 5. Final thoughts

DNA computing is an exciting line of recent research. The advantage with this is that the processors (the DNA molecules) are of microscopic sizes and thus offer great possibilities for parallelism. Quantum mechanical effects also only occur on a microscopic scale. In case it is possible to design parallel quantum mechanical algorithms so that each processor follows identical rules, it presents the same possibility for large scale parallelism, this is in addition to the inbuilt parallelism of quantum mechanical effects that existing quantum mechanical algorithms use [Factor94], [Search96], [Median96]. This paper gives such an algorithm for the problem of estimating the mean where the speed-up using $\eta$ processors is $O(\eta)$.

NMR computing [NMR97a] [NMR97b] are recent implementations of parallel quantum computing. As presently implemented, it does not assume coherence between different molecules. It is equivalent to classically combining the results from various quantum computers - to use another analogy it is like incoherently combining the radiation from sources that are intrinsically coherent (e.g. a light bulb as opposed to a laser).

## 6. Acknowledgments

Would particularly like to thank Artur Ekert for pointing out the [Freq96] result, John Preskill for pointing out an inconsistency in the problem formulation, Norm Margolus for critically reviewing the manuscript & Gilles Brassard, whose seminar on quantum teleportation inspired some of the ideas in this paper. Discussions with Andre Berthiaume, Andy Yao, Bernie Yurke, Chris Monroe & Tad Hogg were helpful.

## 7. References

[Factor94]   P. Shor, *Algorithms for quantum computation: discrete logarithms and factoring,* Proceedings, 35th Annual Symposium on Fundamentals of Computer Science (FOCS), 1994, pp. 124-134.
[Freq96]   J. J. Bollinger, W. M. Itano, D. J. Wineland & D. J. Heinzen, *Optimal frequency measurements with maximally correlated states,* Physical Review A, Vol. 54, No. 6, Dec. 1996.
[Median96]   L. K. Grover, *A fast quantum mechanical algorithm for estimating the median,* Bell Labs Technical Memorandum ITD-96-30115J (lanl e-print quant-ph/9607024).
[Model93]   E. Bernstein & U. Vazirani, *Quantum complexity theory,* Proceedings, 25th ACM Symposium on Theory of Computing (STOC), 1993, pp. 11-20.
[NMR97a]   N. Gershenfeld and I. L. Chuang, *Bulk spin resonance quantum computation,* Science, vol. 275, no. 5298, page 350, 1997.
[NMR97b]   D.G. Cory, A.F. Fahmy & T.F. Havel, *Ensemble quantum computing by NMR spectroscopy,* Proceed-

[----]  ings, National Academy of Sciences, pp. 1634-1639.

[Qntcmm97]  R. Cleve & H. Buhrman, *Substituting quantum entanglement for communication,* (lanl e-print quant-ph/9704026).

[Search96]  L. K. Grover, *A fast quantum mechanical algorithm for database search,* Proceedings, 28th ACM-Symposium on Theory of Computing (STOC), 1996, pp. 212-218, (lanl e-print quant-ph/9605043).

[Teleport93]  Bennett et al, *Teleporting an unknown quantum state via dual classical & EPR channels,* Phys. Rev. Letters, Vol. 70, no. 13, March 29, 1993, pp. 1895-1900.

[Transform92]  D. Deutsch & R. Josza, *Rapid solution of problems by quantum computation,* Proc. Royal Society of London, A400, 1992, pp. 73-90.